\documentclass{article}
 \usepackage[english]{babel}
 \usepackage[affil-it]{authblk}
\usepackage[utf8]{inputenc}
\usepackage{booktabs} 
\usepackage{adjustbox}
\usepackage{amsmath,amssymb, bm, bbm,empheq}
\usepackage{stmaryrd}
\usepackage{csquotes}
\usepackage{multicol}
\usepackage{graphicx}
\usepackage{subfig}
\usepackage[pdftex]{hyperref}
\usepackage{algorithm}
\usepackage{algorithmic}

\usepackage[natbib=true, style=numeric, sorting=nyt]{biblatex}
\providecommand{\keywords}[1]{\textbf{\textit{Keywords --}} #1}
\DeclareUnicodeCharacter{2212}{-}
\newcommand{\ra}[1]{\renewcommand{\arraystretch}{#1}} 

\newcommand{\ReLU}{\text{ReLU}} 
\newcommand{\acc}{\text{acc}} 
\newcommand{\conf}{\text{conf}} 

\title{Credit scoring using neural networks and SURE posterior probability calibration \footnote{This work was supported by the Caisse des dépôts et consignations}
}

\author[1]{Matthieu Garcin}
\author[1,2]{Samuel Stéphan \thanks{Corresponding author}}
\affil[1]{Léonard de Vinci Pôle Universitaire, Research center, 92916 Paris La Défense, France.}
\affil[2]{SAMM, Université Paris 1 Panthéon-Sorbonne, 90 rue de Tolbiac, 75013 Paris cedex 13, France.}

\begin{document}

\maketitle



\begin{abstract}
In this article we compare the performances of a logistic regression and a feed forward neural network for credit scoring purposes. Our results show that the logistic regression gives quite good results on the dataset and the neural network can improve a little the performance. We also consider different sets of features in order to assess their importance in terms of prediction accuracy. We found that temporal features (i.e. repeated measures over time) can be an important source of information resulting in an increase in the overall model accuracy. Finally, we introduce a new technique for the calibration of predicted probabilities based on
Stein's unbiased risk estimate (SURE). This calibration technique can be applied to very general calibration functions. In particular, we detail this method for the sigmoid function as well as for the Kumaraswamy function, which includes the identity as a particular case. We show that stacking the SURE calibration technique with the classical Platt method can
improve the calibration of predicted probabilities.
\end{abstract}

\keywords{Deep learning, credit scoring, calibration, SURE}

\section{Introduction} \label{intro}

Credit scoring aims to measure the risk for a bank to grant a loan to an applicant. Depending on the value of the returned score, the bank will accept or not to grant the loan. This score is generally computed by a model which has been fitted on a database containing past information of the consumer behavior and its corresponding credit profile. Typical descriptors of consumer behavior include loan information (amount, maturity, type of interest rate, nature of the loan) and the borrower characteristics (age, marital status, profession, monthly income, personal savings, number of actual loans). The target, also known as dependent feature, corresponds to whether or not the customer has defaulted on its loan. The default is subject to a formal definition given by the Basel committee which states that the bank is facing a default event if the counterpart past due is more than 90 days. Thus, the feature to be predicted is coded as 1 if the borrower did default and 0 otherwise. 

Correct risk assessment is an important aspect of banking activities. To ensure robust estimation of the risks, the regulator has framed several rules to follow. These rules are edited by the Basel Committee on Banking Supervision (BCBS) which is the primary global standard setter for the prudential regulation of banks. Several updates have been made by the regulator to fit the development of banking activities over the years. Currently, banks are subject to Basel III agreements that aim to ensure that banks have a rigorous approach to risk and capital management related to their activities. The Basel framework lets the choice for each financial institution to manage its credit risk assessment through standard or internal methods. In the latter case, the bank uses a modeling framework for estimating the risk parameters. In the Internal Rating Based (IRB) approach, the bank estimates only the probabilities of default (PD). In the advanced IRB, the bank has to estimate the expected credit loss using the following parameters: the Loss Given Default (LGD), Exposure at Default (EAD), Maturity of exposures (M), and the PDs. As its name indicates, the PD risk parameter consists in estimating the likelihood that each loan is going to be repaid. This is typically what is evaluated in credit scoring applications. 


From a practical point of view, the estimation of the PDs corresponds to a classification problem. The specificity of this problem is that we want the classifier to output a probability of the event while in numerous other applications we only want the outcome. Traditional machine learning models such as logistic regression and linear discriminant analysis are well suited for this task~\cite{Steenackers89,Altman68}. Despite their simplicity, they are probably the most widespread models for credit scoring applications since they are very well understood tools and easy to use~\cite{Anderson07,Dumitrescu20}. Furthermore, they are implemented in most statistical software and the computation of the final prediction is straightforward from the features coefficients. Today, a large area of research in credit scoring consists in the development of new scoring techniques. This literature is motivated by the limitations of the standard techniques. Indeed, in their simplest design, logistic regression and linear discriminant analysis exploit only linear interactions.
Thus, the rise of performances of credit scoring models has been initiated by the use of ensemble methods that enable non-linearities and provide high generalization capabilities~\cite{Baesens03,Finlay11,Paleologo10,Wang11}. Other recent advances include the use of hybrid methods~\cite{hybrid_svm,hybrid_neural,hybrid_deep} and deep learning which have first achieved promising results in the field of computer vision~\cite{LeCun98,LeCun16,Voulodimos18,neural_CS_1,neural_CS_2,neural_CS_3}. Nevertheless, credit scoring remains a field where deep learning has trouble asserting itself because of strong regulatory requirements. 

In this paper, we investigate how powerful are these deep neural networks through a feed-forward architecture and we compare the results with a standard logistic regression. We show that the use of a deep neural network leads, on our dataset, to a slight improvement in the forecast compared to the logistic regression. We also investigate the properties of our dataset which has the particularity to contain a mix of static features and temporal features. To the best of our knowledge, no previous study on credit scoring has ever explored the ability of a model to provide accurate results depending on the static/dynamic nature of the data. The procedure we use is as follows: we split the dataset into static features set and dynamics features set. We then evaluate the models on three sets: static, dynamic, and all features, and finally compare the results. We give evidence that temporal data should not be neglected in credit scoring applications. In our dataset, dynamic features drastically improved the results compare to feeding the model only with static features. 
 
Another important aspect of credit scoring application is to ensure that the estimated probabilities are close to the true probabilities. This can be analyzed in terms of cost. On one hand, machine learning models assume all estimated probabilities to have the same cost. On the other hand, in many decision-making applications, not having an accurate probability of belonging to the target class can be costly. For instance, it can be the choice of a treatment for a patient in medicine, a driving decision in a self-driving car application, or an investment decision in business. A common approach consists in post-processing the probabilities of a classifier in order to get calibrated outputs. Calibration can be performed using parametric and non-parametric techniques. Non-parametric techniques include histogram binning~\cite{Zadrozny01}, isotonic regression~\cite{Zadrozny02}, similarity binning averaging~\cite{Bella}, and adaptive calibration of predictions~\cite{Jiang12}. All these techniques consist in binning the samples and assigning them a calibrated probability. We argue that probabilities calibrated by these techniques can be highly biased. Indeed, each bin requires a big enough number of instances in order to have a low variance estimate of its average default rate, thus leading to a unique estimated PD for all the instances in this bin. Choosing the optimal number of bins, balancing bias and variance, is not obvious as well. Parametric methods include Platt scaling~\cite{Platt99}, beta calibration~\cite{Kull17}, asymmetric Laplace method~\cite{Bennett03}, and piecewise logistic regression~\cite{Zhang04}. Since these parametric approaches rely on the assumption that the probabilities follow a particular distribution, they are subject to model mismatch. In this paper, we propose a novel parametric approach for probability calibration. We use Stein's Unbiased Risk Estimate (SURE) as a proxy to minimize the Mean Squared Error (MSE) between the estimated probability and the true probability. The calibrated density is then the one minimizing the estimated MSE. This technique has been used intensively in the field of signal processing for image denoising. One advantage of this approach is that we do not rely on predefined bins which are biased by nature. Besides, our method offer the possibility to use a custom calibration function to prevent from model mismatch. We show empirically that combining this new technique with the standard Platt method can result in an increase in the accuracy of the PDs. 

The major contribution of this article to the credit scoring literature is the investigation of the efficiency of models based on deep learning versus traditional models when estimating individual PDs. We also evaluate the impact of using time series data in credit scoring models. Finally, we propose a methodology to calibrate the predicted probability to the hidden true probabilities using the SURE approach. 
The article is organized as follows. Section~\ref{expsetting} describes the data used for the application, exposes the evaluation of the models and  the feature importance assessment, and presents the models. In Section~\ref{calibrationframework}, we demonstrate the necessity of calibration in machine learning and we propose a new framework to calibrate posterior predicted probabilities. Section~\ref{results} compares empirically the models and assesses the efficiency of the proposed calibration approach. 

\section{Experimental setting} \label{expsetting}

In this section, we describe how we design the experiment. First, we present the dataset on which models have been fitted. We precise what preprocessing steps have been applied and how we split the data in order to assess the performance of our model. Next, we explain how we organized the features such that we were able to assess the importance of static and dynamic features in the performance. Then, we discuss the choice of a proper evaluation metric in credit scoring. Indeed, credit scoring is an area in which the event to predict is difficult because of class imbalance. This characteristic should drive the choice of the measure to get an unbiased measure of performance. Finally, we present the models we use for the application. 

\subsection{The dataset} \label{data}

In every machine learning application, the quality of data is of primary importance. Ideally, a credit scoring application would include quantitative data such as financial ratios describing the borrower's financial health and its past credit history, including potential default information. Indeed, defaults have been shown to be persistent over time with the reasoning that a borrower who has already defaulted is more likely to stay in this current state~\cite{neural_CS_2}. Qualitative data are useful as well, such as the education level, the type of product granted, or the presence of collateral to secure the loan. 

In this study, we use a Taiwanese credit dataset publicly available on the UCI machine learning repository\footnote{https://archive.ics.uci.edu/ml/datasets/default+of+credit+card+clients} and already used in the machine learning literature~\cite{creditscoring1}. This dataset consists of anonymized default payments realized on revolving credit granted in Taiwan. It gathers information of 30 000 credit card users among which 22\% default next month. We can notice that the dataset is imbalanced between defaulters and non-defaulters. This is a common characteristic of credit scoring applications which makes the modeling difficult. The dataset includes classic features of a credit scoring dataset such as age, sex, level of education, marital status, and the credit limit of individuals. It also contains time series data observed from April 2005 to September 2005: the monthly bill statement, the monthly paid amount, and an indicator of the revolving credit delay payment. We expect these last dynamic features to be highly informative since it gives a trajectory of the individual account which may reveal interesting patterns for defaulters. Our target is the dummy variable indicating whether the client defaults next month (No = 0, Yes = 1). To complete the dataset, we have created 6 additional features, one for each month of observation, consisting of the ratio of the bill amount by the credit limit. We expect this set of new features to improve the model's ability to learn the default drivers, since a high bill statement relative to the credit limit may imply payment difficulties. 
 
We apply the classical preprocessing steps to the features in order to facilitate training. We standardize the distribution of continuous data to have a mean value of 0 and a standard deviation of 1. This typical operation is done in order to speed up the learning process. We then split the dataset into a shuffled and stratified train, a validation, and a test set representing respectively 60\% - 20\% - 20\% of the data available. 
We use the train and validation sets to train each model and tune their hyperparameters while we use the test set only once to assess the out-of-sample performance. This last set allows us to estimate an unbiased performance of the classifier. 

\subsection{Feature importance assessment}

In this study, we are interested in assessing the importance of different sets of features instead of each feature individually. Indeed, we want to assess the importance of the nature of data in the model's performance. In particular, we expect times series to be highly informative since they reflect the behavior of the borrower. Thus, we defined three subsamples of features which are detailed in Table~\ref{tab:features}. Note that all dynamic data are time series available from April 2005 to September 2005. 

\begin{table*}[h!]
\ra{1.5}
\begin{tabular}{p{0.25\linewidth} p{0.75\linewidth}}
\toprule
Static features & Amount of given credit (NT dollar), Gender, Marital status, Age (year). \\ 
Dynamic features & History of past payment tracked via past monthly payment records, Repayment status, Amount on bill statement, Amount of previous payment, Ratio of bill statement relative to the credit limit \\
All features & Static features + Dynamic features \\
\bottomrule
\end{tabular}
\caption{Features sets}
\label{tab:features}
\end{table*}

We expect time series data to be highly informative since it gives a dynamic view of the borrower's credit activity. Moreover, classical scoring models often incorporate an aggregated vision of the borrower's behavior instead of a detailed monthly vision. We think that the industry could benefit from the use of time series data in their internal credit scoring models in terms of predictive accuracy. Using this particular setting, we are able to train each model on different sets of features and evaluate their inner performances on the corresponding independent testing sets. Then, we are able to identify the most informative set of features by picking the one that achieves the highest performance on the test set.

\subsection{Choosing the right evaluation metric} \label{evaluationMetric}

Credit scoring is typically an area where class imbalance can be severe because risk management aims to ensure that the number of defaulters stays much smaller than non-defaulters. Such characteristics make learning difficult. This is why several techniques have been introduced. Data-level techniques consist in resampling the dataset in order to balance the classes. These techniques include random over-sampling, random under-sampling, and the creation of synthetic samples via the SMOTE algorithm~\cite{He09,Chawla02}. Other strategies for balancing the classes consist in rendering the model cost-sensitive by adding a cost to misclassified instances of both positive and negative class. This improves the model's ability to correctly classify positive samples. Such methods include weighting and thresholding~\cite{Sun09,Sheng06}. In this article, our choice goes to weighting the loss function as detailed in section~\ref{models:LR}. This choice is motivated by the fact that being a model-level technique, it doesn't change the structure of the data. Therefore, the original statistical properties of the dataset are preserved which is preferable from a regulatory point of view. We also applied thresholding to the predicted probabilities as explained below. 

The class imbalance also makes the choice of the appropriate evaluation metric quite challenging. Indeed, standard measures can't be applied because they tend to be biased toward the majority class. Among these evaluation metrics, we can cite, for instance, the accuracy, which is defined as the sum of true positives and true negatives over the whole dataset. The accuracy is not a good choice of metric since a classifier biased toward the negative class will always reach a high accuracy. In this study, we propose an evaluation based on the precision, the recall and the F1 score. Let's consider a classical binary classification problem with a model fitted on a training set. We want to evaluate this model on a test set. We define $y_i \in \{0,1\}$ as the true label of the $i^{th}$ instance. The label $y_i$ equals 1 if the instance is tagged as positive and 0 if tagged as negative. The model outputs a quantity $\widehat{p_i} \in \left[0,1\right]$ which is often interpreted in the literature as the probability of a given instance of belonging to the positive class. For a given threshold $\tau \in \left[0,1\right]$, the predicted label is defined as $\widehat{y_i}=1$ if $\widehat{p_i}> \tau$ and $\widehat{y_i}=0$ otherwise. 
Given these notations, we can compute the number of True Positives ($TP=\sum_{i=1}^{N} \mathbbm{1}_{(y_i=1 \cap \widehat{y_i}=1)}$), True Negatives ($TN=\sum_{i=1}^{N} \mathbbm{1}_{(y_i=0 \cap \widehat{y_i}=0)}$), False Positives ($FP=\sum_{i=1}^{N} \mathbbm{1}_{(y_i=0 \cap \widehat{y_i}=1)}$), and False Negatives ($FN=\sum_{i=1}^{N} \mathbbm{1}_{(y_i=1 \cap \widehat{y_i}=0)}$). Then we can compute the recall, the precision, and the F1 score:
$$\text{Recall} = \frac{\text{TP}}{\text{TP} + \text{FN}}$$ 
$$\text{Precision} = \frac{\text{TP}}{\text{TP} + \text{FP}} $$
$$
 \text{F1} = 2 \times \frac{\text{Precision} \times \text{Recall}}{\text{Precision} + \text{Recall}}. 
$$
 
The F1 score is a combination of the precision and the recall. Maximizing this metric thus leads to balance the effect of false negative and false positive, giving an unbiased estimate of how the model is performing. 

Another well-known measure for model evaluation is the ROC curve. This curve is obtained by computing the specificity and the corresponding sensitivity for various probability thresholds of the classifier. We then plot the sensitivity against the specificity. One convenient measure associated with the ROC curve is the area under the curve (AUC), with AUC $> 0.5$ meaning that the model is performing better than a random classification. These metrics are also not good candidates for the evaluation of imbalanced datasets classifiers. The reason is that the ROC curve is used to assess the overall performance in discriminating the positive class and the negative class while most of the time, in imbalanced learning, we are only focusing on the correct classification of positive samples. This generally leads to an overconfident estimate of how well the classifier is performing because it doesn't take into account that the classifier is biased toward the negative class containing more samples. Instead, we can rely on the Precision-Recall (PR) curve which focuses on the performance of the model in classifying positive samples only~\cite{Davis2006,precisionRecall}. This curve is obtained by plotting the precision and recall for various probability thresholds. As in the ROC case, we can compute the corresponding AUC of the PR curve~\cite{Boyd13}. 
 
We consider the recall, the precision and, the F1 score to evaluate the performance of our classifiers for two reasons. First, our dataset is imbalanced as it contains 22\% of defaulters. Second, in this imbalanced framework, our preference goes to a better classification of positive samples, in order to diminish the risk for the lender. These measures being based on the confusion matrix, the attribution of classes is ultimately done by applying a threshold value to all samples. Most software considers a default probability threshold of 0.5 which is not recommended for imbalanced datasets~\cite{Provost,optThreshold, optThreshold2}. We choose to tune the threshold value such that we achieve the best F1 score in a simple manner. The procedure is as follows: we move the decision threshold for the predicted probabilities of the training set. For each of these thresholds, we assign a class to each sample and compute the corresponding F1 score. We then apply this threshold to the test set to compute the out-of-sample F1 score. We also consider a global measure of performance such as the AUC-PR and display the AUC-ROC for comparison purposes.

\subsection{Models} \label{models}
\subsubsection{Logistic regression} \label{models:LR}

The logistic regression will be our baseline model since it is still one of the most widely used in the banking industry for credit scoring~\cite{Anderson07}. To take into account the slight imbalance of classes (i.e. 22/78), we trained the model using the balanced cross-entropy loss~\cite{Xie15}. We consider the learning set $\{(x_i,y_i) \in (\mathcal{X} \times \mathcal{Y}) | i=1,...,N\}$ where $y_i$ are the labels, $x_i$ the inputs, and $N$ the total number of instances. We want to estimate a function $f(x|\theta)$ which maps the inputs $x_i$ to the output $y_i$, where $\theta$ is a set of parameters to be optimized on the training set. We denote $\widehat{p_i} = f(x|\theta)$ the model's output and $n^{+}$ the number of positives samples. We estimate $\theta$ by minimizing the loss function over all the instances:


\begin{equation}
 \mathcal{L}(y_i,\widehat{p_i}) = - [\alpha y_i\log(\widehat{p_i}) + (1-\alpha)(1-y_i)\log(1-\widehat{p_i})], 
\label{loss}
\end{equation}

$\text{where } \alpha=\frac{n^{+}}{\text{N}}.$ Intuitively, the term $\alpha$ accounts for class imbalance insofar as mispredictions on both the positive and the negative class are penalized.

\subsubsection{Feed forward neural network}\label{models:FFNN}

We specify a neural network with a feed forward architecture for predicting the occurrence of defaults. The advantage of such a model over traditional methods is that we can easily modify the network in order to perfectly scale the problem. This is stated in the universal approximation theorem~\cite{Hornik91}: \begin{displayquote}
"A single hidden layer neural network with a linear output unit
can approximate any continuous function arbitrarily well,
given enough hidden units."
\end{displayquote}
Moreover, it has been shown that it is more efficient in terms of predictive performance to build a multi-layer neural network since it is able to learn deeper representations~\cite{Bengio07}. Convolutional neural networks for instance are able to learn complex concepts such as edges or geometric shapes of an image. Also, multi-layer neural networks need exponentially fewer neurons than shallow networks to learn data representations. 

Our network architecture is composed of an input layer, three hidden layers of 60 hidden units each, and an output unit which uses a sigmoid activation function in order to output a probability estimate. We introduce nonlinearities in our network with the help of the ReLU activation function defined by: $$\ReLU(x) = \max(0,x).$$

This particular choice of activation function is motivated by the literature because it decreases the training time and can approximate any continuous function ~\cite{Krizhevsky12}. We train the network using the balanced cross-entropy loss presented in equation \eqref{loss}. This common framework makes it possible to compare fairly the performance of the two models since their prediction abilities will come from their intrinsic architecture. 

The weights of the network have been optimized using the Adam algorithm with a batch size of 256 samples~\cite{Adam}. Based on the literature, we initialize carefully the weights using He initialization scheme~\cite{Bengio10,He}. This approach is widely used by practitioners to ensure that gradients don't vanish or explode during training. 


\section{Probability calibration} \label{calibrationframework}

We first review the existing literature on calibration. We present the original use of SURE and we connect this approach to the machine learning area. Then, we detail our SURE calibration framework and we give a pseudo-code to enhance the comprehension and allow reproducibility. We conclude this section by giving evidence of why our framework should be preferred to binning and we report the evaluation measure we used to estimate the calibration error. 

\subsection{Literature review}
Ideally, we would like machine learning models to output accurate probabilities in the sense that they reflect the real unobserved probabilities. This is exactly the purpose of calibration techniques, which aim to map the predicted probabilities to the true ones in order to reduce the probability distribution error of the model. Probability calibration is very important in many real-world classification tasks. Indeed, most of the time, classification models are evaluated globally, using synthetic measures, without taking care of how the error is distributed. Thus, one can't identify the confidence of the predicted probabilities. Calibration is overriding for instance in medicine~\cite{CalMedecine,CalMedecine2}, for self-driving cars~\cite{CalDriving}, or in finance~\cite{CalFinance}. Indeed, reliable probabilities are preferable to take accurate decisions and reduce the risks.

In credit scoring applications, having accurate probabilities is a matter of concern in the perspective of improving risk assessment. Good calibration can result in significant gains for financial institutions since they will be able to correctly assess the risk related to each borrower~\cite{CalBenefits}. Besides, predicted probabilities that correctly match the empirical distribution lead to better risk management and anticipation when it comes to model different portfolio scenarios and evaluate expected losses~\cite{CalFinance}. 

Usually, calibration of predicted probabilities is performed using histogram binning ~\cite{Zadrozny01}, Platt scaling~\cite{Platt99}, or isotonic regression~\cite{Zadrozny02}. Concretely, histogram binning consists in dividing the samples into equal bins and assigning them a calibrated probability. In the case of Platt scaling and isotonic regression, the method consists in fitting a model to regress the predicted probabilities on the real labels. The calibration model is estimated on the validation set and is then used to predict default probabilities on the test set. 
Recently, calibration has been rediscovered with the deep learning booming trend. Indeed, deep learning models are no exception to the rule and are even more than traditional methods subject to model uncertainty due to their complex architectures~\cite{CalDeepNet}. That's why measuring calibration error and develop calibration techniques for deep neural networks have become an important research topic these recent years~\cite{MeasureCal,tempscalingNN,FieldCal}. 

In this paper, we are introducing a new calibration method based on SURE. This method has originally been developed to estimate the MSE\footnote{Called risk in the literature about signal processing, without any link with the finance-related notion of risk introduced in this paper.} of any nonlinear differentiable estimator whose input observations are assumed to be Gaussian~\cite{Stein}. We think that this approach is well suited for probabilities calibration since it allows to measure an unobserved error. In our case, this error is the difference between the real unobserved probability and the estimated one. This technique has been used intensively in signal processing, mostly as a denoising technique. One of the major applications in this field was to select the optimal threshold of noisy wavelet coefficients in order to recover a signal~\cite{SignalDenoising,WaveletShrinkage}. The SURE method also appears in the machine learning literature for model selection. Many modern machine learning algorithms require shrinkage through a regularization parameter to get models that generalize well to new data. In this case, minimizing the SURE can be an approach to find this optimal regularization parameter~\cite{RegSURE,Efron04,Zou07,Tibshirani11,Zhu08}. Currently, most practitioners use cross-validation as an estimate of the model's MSE because it doesn't require normality assumptions and it's easy to implement. However, research has shown that both methods give robust results~\cite{RegSURE}. 

In the deep learning area, the SURE method is used for neural network based denoising algorithms. It has been proposed to train a deep denoiser network only on noisy training~\cite{DeepDenoiser18}. The model outperformed the classical non-deep learning based denoisers. This approach has been extended for learning with correlated pairs of noisy images and compressed sensing~\cite{DeepDenoiser19,UnsupervisedStein}.

In this work, we calibrate the output of our models using our SURE framework. This choice is motivated by the fact that we want to minimize the error depending on the unobserved true probabilities. Thus cross-validation cannot be applied since we never observe the true probabilities. Furthermore, we argue that our SURE framework gives a better estimation of the calibration error than histogram binning. Indeed, in histogram binning, the choice of the number of bins introduces a bias due to the unequal number of samples falling in each bin. Finally, we compare our method to Platt scaling which is a parametric method in the same vein as our SURE framework. 

\subsection{Platt scaling}\label{Plattscaling}

In this subsection, we give a brief overview of the Platt scaling method which is the standard approach for probabilities calibration. This approach consists in using the predicted probabilities of a classifier as an input of a logistic regression. Thus, probabilities are modified by the sigmoid function below: 
\begin{equation}
  \widetilde{p_i} = \frac{1}{1+e^{-(\theta_1\widehat{p_i}+\theta_2)}},
  \label{eq:sigmoid}
\end{equation}

where $\widehat{p_i}$ is the predicted probability of observation $i$, $\widetilde{p_i}$ is the calibrated probability and $\theta_1$, $\theta_2$ the parameters of the logistic regression. These parameters are to be estimated so as to make the calibrated probability close to the true probability $p_i$. However, $p_i$ is not observed and Platt's method puts forward the labels $y_i$ instead. Thus, we estimate $\theta_1$ and $\theta_2$ by maximizing the likelihood, in the calibrated probability, of observed data, namely the labels. The labels can take values 0 or 1.  Thus, it can be interpreted as the realisation of a Bernoulli random variable $Y$ such that $Y \sim Ber(\widetilde{p_i})$. Given this general framework, we can write the likelihood of the observations:

$$L(\theta_1,\theta_2) = \prod_{i=1}^{N} \left( \frac{1}{1+e^{-(\theta_1\widehat{p_i}+\theta_2)}}\right)^{y_i} \left(1-\frac{1}{1+e^{-(\theta_1\widehat{p_i}+\theta_2)}} \right)^{1-y_i}.$$

This equation is often transformed in log-likelihood for calculus convenience:

$$\mathcal{L}(\theta_1,\theta_2)= \sum_{i=1}^{N} y_i \log\left(\frac{1}{1+e^{-(\theta_1\widehat{p_i}+\theta_2)}}\right) + (1-y_i)\log\left(1-\frac{1}{1+e^{-(\theta_1\widehat{p_i}+\theta_2)}}\right).$$

The log-likelihood is also known as Binary Cross Entropy (BCE). We can maximize this quantity by applying gradient-based algorithms such as steepest descent or quasi-Newton to find an estimate of the parameters $\theta_1$ and $\theta_2$.

\subsection{A SURE framework for calibration}

\subsubsection{General framework}
In this framework, we consider the predicted probabilities $\bm{\widehat{p}} = (\widehat{p_i})_{i\in \llbracket1,N \rrbracket}$ of a model to be noisy measurements of the true probabilities. We make the assumption that the relation between the corrupted probabilities and the true probabilities is as follows:
$$\bm{\widehat{p}} = \bm{p} + \bm{\varepsilon},$$
where $\bm{p} = (p_i)_{i\in \llbracket1,N\rrbracket}$ denotes the true probabilities and $\bm{\varepsilon} = (\varepsilon_i)_{i\in \llbracket1,N \rrbracket}$, the noise, is a vector of i.i.d. Gaussian variables of mean 0 and variance $\sigma^2$. 

We also define the denoising function $G_{\theta}(.)$ parametrized by $\bm{\theta}$ and weakly differentiable. This function takes in input the noisy probabilities and output the denoised probabilities that is $\widetilde{p_i} = G_{\theta}(\widehat{p_i})$. Besides, we restrict the number of parameters to two, that is $\bm{\theta}=(\theta_1,\theta_2)$. This seems reasonable since Platt method also consists in two parameters. The function $G_{\theta}(.)$ is increasing, so that the order of the denoised probabilities is respected: if $\widehat{p_i}<\widehat{p_j}$, the calibration must not change the classification and thus $\widetilde{p_i}<\widetilde{p_j}
$.

Our goal is to select optimally the parameters $\bm{\theta}$ in order to obtain an estimate $\bm{\widetilde{p}} = (\widetilde{p_i})_{i\in \llbracket1,N \rrbracket}$ as close as possible to $\bm{p}$. This kind of problem is typically achieved by minimizing the MSE between the estimated probability and the true probability:
\begin{equation*}
 r(\bm{\widetilde{p}}) = \frac{1}{N}\sum_{i=1}^{N}(\widetilde{p_i} - p_i)^{2}. 
\end{equation*}

However, in practice, we don't have access to $\bm{p}$. The SURE method proposed by Stein~\cite{Stein} allows us to overcome this difficulty by providing an approximation of the MSE, defined by:

\begin{equation}
 SURE(\bm{\theta}) = -N\sigma^{2} + \sum_{i=1}^{N}(G_{\theta}(\widehat{p_i}) - \widehat{p_i})^2 + 2\sigma^{2}\sum_{i=1}^{N} \frac{\partial G_{\theta}(\widehat{p_i})}{\partial \widehat{p_i}}.
 \label{eq:sureloss}
\end{equation}

The  SURE  loss  is  an  unbiased  estimate  of  the  MSE  under  the  assumption  of  Gaussian  noise. Since we want an unbiased estimator, we expect the mean calibrated probability to be equal to the empirical event frequency that is $\frac{1}{N}\sum_{i=1}^{N}G_{\theta}(\widehat{p_i}) = \frac{1}{N}\sum_{i=1}^{N}y_i$. This leads to the following minimization program:

\begin{empheq}[left={(\mathcal{P})=\empheqlbrace}]{align*}
\min_{\bm{\theta}} \quad &\text{SURE($\bm{\theta}$)}\\
\textrm{s.c.} \quad &C(\bm{\theta})=0
\end{empheq}

with $C(\bm{\theta})=\frac{1}{N}\sum_{i=1}^{N}G_{\theta}(\widehat{p_i})-\frac{1}{N}\sum_{i=1}^{N}y_{i}$. 

\subsubsection{Choice of the calibration function}

In this experiment, we tried the sigmoid and Kumaraswamy repartition functions as denoisers. Both are increasing and differentiable functions on the interval $[0,1]$. This properties make them suitable for the calibration of probabilities. 

The sigmoid function defined in equation~(\ref{eq:sigmoid}) is the one we used in the experiment. This choice seems natural since it is also the one used in Platt scaling. Thus, it makes sense to use it for fair comparison of both methods. This distribution has the particularity to assume normally distributed probabilities within each class with same variance. This strong property can reduce the efficiency of the approach which motivates the choice of another function~\cite{Kull17}.

The Kumaraswamy distribution has been presented first for simulations purposes in hydrological data modeling ~\cite{Kumaraswamy}. This function is of the same family of the beta law and it has the advantage to be explicitly differentiable. We define the cumulative Kumaraswamy distribution function as:

$$G_{\theta}(\widehat{p_i})= 1-(1-\widehat{p_i}^{\theta_1})^{\theta_2},$$

with $\theta_1>0, \theta_2>0.$

This function is much more flexible than the sigmoid one because it can map the predicted probabilities to the sigmoid, the inverse sigmoid, and the identity shapes. In particular, if probabilities are already calibrated, the Kumaraswamy can recover the identity while the sigmoid does not, independently to the method employed. The ability to recover different shapes makes the Kumaraswamy suitable for a large variety of models that output different uncalibrated probability distributions.  

\subsubsection{Estimation of the noise variance}

In the above framework we assumed the knowledge of $\sigma^2$ which is not true in practice. We want to estimate $\sigma$ from the data such that it stays constant over the optimization. We have made the assumption that $\widehat{p_i}=p_i + \varepsilon_i$, with $\varepsilon_i \sim N(0,\sigma^{2})$. Besides, the labels $y_i$ can be equal to 0 or 1. Thus each observation $y_i$ can be interpreted as the realisation of a Bernoulli random variable such that $y_i \sim Ber(p_i)$. Hence we have $\mathbb{P}(y_i=1) = p_i$ and $\mathbb{P}(y_i=0) =1-p_i$.

We want to minimize the spread between the estimated probability $\widehat{p_i}$ and the true one $p_i$. Thus, taking the expectation:  

$$ \mathbb{E}[(\widehat{p_i} - y_i)^{2}] =\mathbb{E}[\widehat{p_i}^{2}] +\mathbb{E}[y_i^{2}] - 2\mathbb{E}[\widehat{p_i}y_i],$$
  
where: 
\begin{align*}
\mathbb{E}[\widehat{p_i}^{2}] &=\mathbb{E}[(p_i+\varepsilon_i)^{2}] = p_i^2 + \sigma^2  \\ 
\mathbb{E}[y_i^{2}] &= 0^2\mathbb{P}(y_i=0) + 1^2\mathbb{P}(y_i=1)=p_i \\
\mathbb{E}[\widehat{p_i}y_i] &=\mathbb{E}[(p_i+\varepsilon_i)y_i]=\mathbb{E}[(p_i+\varepsilon_i)]\mathbb{E}[y_i] = p_i^{2}.
\end{align*}

In the last calculus, we assume the independence between $p_i$ and $y_i$ to conclude, which is obviously not guaranteed. However, this strong assumption gives a simple expression of $\sigma^{2}$ and leads to satisfying empirical results. 

We end up with: 
$$ \mathbb{E}[(\widehat{p_i} - y_i)^{2}] =p_i^2 + \sigma^2 + p_i -2p_i^{2} = p_i(1-p_i) + \sigma^{2} = V(y_i) + \sigma^{2}.$$
  
Rearranging the terms, we have the theoretical expression for $\sigma^{2}$: $\sigma^{2}=\mathbb{E}[(\widehat{p_i} - y_i)^{2}] - V(y_i) $, which can be estimated by the following equation:
$$\widehat{\sigma}^{2}=\frac{1}{N}\sum_{i=1}^{N}(\widehat{p_i} - y_i)^{2} - \frac{1}{N}\sum_{i=1}^{N}(y_i - \overline{y})^{2}, $$

where $\overline{y}$ is the mean of the labels $y_i$. 

We also note that, alternatively, $\sigma^{2}$ can be cross-validated in order to optimize a criteria. This technique has the advantage of giving a robust estimate of $\sigma$ that will generalize well on the unseen data. We made several trials that suggests that using the sigmoid function as a denoiser, we can retrieve exactly the Platt calibration map by cross-validating $\sigma^{2}$.  

\subsubsection{Numerical approximation framework}
To solve the constrained optimization problem $\mathcal{P}$, we can reformulate it as a problem of minimization of a function $\mathcal{Q}$ composed of the original objective function and of the constraint. Hence, our original constrained problem becomes a sequence of unconstrained optimization problems which can be solved by classical numerical methods. The quadratic penalty function is a natural choice for such a function because of its simplicity. We define it as follows:

$$\mathcal{Q}(\theta,\mu) = SURE(\bm{\theta}) + \frac{\mu}{2} C(\bm{\theta})^{2},$$

where $\mu>0$ is the penalty parameter. This parameter becomes larger if the constraint is not satisfied, such that the overall function is penalized more severely. This forces the minimizer of the penalty function to be close to the feasibility region of the initial problem $\mathcal{P}$.  


One can find a numerical approximation of the solution using first order methods or second order methods since $G_{\theta}(.)$ is a continuously differentiable function. However, even if second order methods are known to be faster than first order ones, these approaches can fail when $\mu$ becomes large because it causes numerical instabilities near to the minimizer~\cite{Optim}. Consequently, we used the steepest descent algorithm to optimize the parameters $\bm{\theta}$. This algorithm requires to compute the gradient which is used as a descent direction. The gradient of our function with respect to the parameters corresponds to the following system:

\begin{empheq}[left=\empheqlbrace]{align*}
\frac{\partial \mathcal{Q}}{\partial \theta_1}=0  \\
\frac{\partial \mathcal{Q}}{\partial \theta_2}=0 
 \end{empheq}
 
that is

\begin{empheq}[left={\empheqlbrace}]{align*}
2\sum_{i=1}^{N}(G_{\theta}(\widehat{p_i})-\widehat{p_i})\frac{\partial G_{\theta}(\widehat{p_i})}{\partial \theta_1} +2\sigma^2 \sum_{i=1}^{N}\frac{\partial G_{\theta}(\widehat{p_i})}{\partial \widehat{p_i} \partial \theta_1} +\mu C(\theta) \frac{\partial C(\theta)}{\partial \theta_1}=0 \\
2\sum_{i=1}^{N}(G_{\theta}(\widehat{p_i})-\widehat{p_i})\frac{\partial G_{\theta}(\widehat{p_i})}{\partial \theta_2} +2\sigma^2 \sum_{i=1}^{N}\frac{\partial G_{\theta}(\widehat{p_i})}{\partial \widehat{p_i} \partial \theta_2} +\mu C(\theta)\frac{\partial C(\theta)}{\partial \theta_2}=0.
 \end{empheq}
 
The explicit calculus of the derivatives for the sigmoid and Kumaraswamy cumulative distribution functions is given in appendix~\ref{ap:sigderive} and \ref{ap:Kumderive}.

The pseudo-code described in Algorithm~\ref{alg:GD_SURE} summarizes the procedure of the proposed SURE calibration.

\begin{algorithm}[H]
\caption{Quadratic penalty method}
 \label{alg:GD_SURE}
\begin{algorithmic}
\STATE Initialization: K= 10, N= 15000, $\theta^{0}$ = (random, random), $\mu^{0}$ = 10, $\alpha$ = 0.0001, tol = 0.00001, $eps$=0.1
\STATE $\widehat{\sigma}^{2}=\frac{1}{N}\sum_{i=1}^{N}(\widehat{p_i} - y_i)^{2} - \frac{1}{N}\sum_{i=1}^{N}(y_i - \overline{y})^{2}$
\WHILE{k$<$K}
\FOR{j = 1 to N}
\IF{$||\nabla_{\theta} SURE|| \leq tol$}
\STATE break;
\ENDIF
\STATE $\theta^{(j+1)} = \theta^{(j)} - \alpha \nabla_{\theta} SURE$
\ENDFOR
\IF{$C(\theta)\leq eps$}
\STATE Save $\theta^{(j)}$
\STATE break;
\ELSE 
\STATE $\mu^{(k+1)} = 10 \times \mu^{(k)}$ 
\ENDIF
\ENDWHILE
\end{algorithmic}
\end{algorithm}

Beyond the SURE calibration technique, we also try a form of model stacking which consists in combining two models in order to improve the overall accuracy. The rationale behind this approach is that each model learns its own non-linearities such that we can benefit from combining models. Here we combine the Platt method to our SURE approach in two ways. We build a first model using Platt's model output probabilities as an input of the SURE model. The second model is the exact reverse: we feed the Platt model by the SURE output probabilities. 

\subsection{Calibration evaluation}

Measuring the calibration error is important to get an estimate of the model uncertainty. In this study, we want to compare the calibrated probabilities of the Platt method to our SURE method. 
Most of the time, calibration models are evaluated using a reliability diagram that compares the average predicted risk (confidence) to the observed proportion of events (accuracy) in each quantile. Following the prevalent notations, we denote the accuracy of the sample $B_m$ as $\acc(B_m)$ and the corresponding average confidence as $\conf(B_m)$~\cite{tempscalingNN}:

\begin{multicols}{2}
\begin{equation*}
 \acc(B_m) = \frac{1}{|B_m|}\sum_{i \in B_m} \mathbbm{1}(\widehat{y_i}=y_i),
\end{equation*} \break 
\begin{equation*}
 \conf(B_m) = \frac{1}{|B_m|}\sum_{i \in B_m} \widehat{p_i},
\end{equation*}
\end{multicols}
 
where M is the total number of bins, $B_m$ is the set of indices of samples whose predicted probabilities falls into the interval $I_m=\mathopen{(}\frac{m−1}{M},\frac{m}{M}\mathclose{]}\text{, for m } \in \llbracket1,…,M\rrbracket$. 

\begin{figure}[htbp]
    \centering
    \includegraphics[width=1\textwidth]{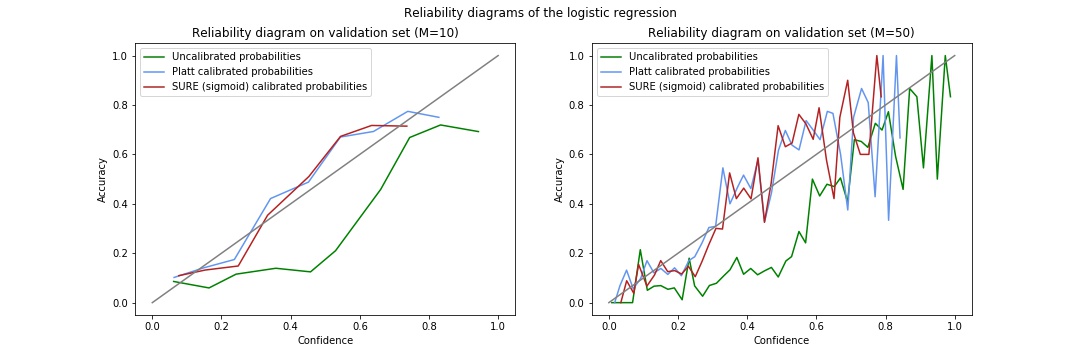}
    \caption{Example of reliability diagrams.}
    \label{fig:RelDiag}
\end{figure}

The key idea of this representation is to visualize the quality of prediction for each quantile of risk. First, predicted probabilities and their corresponding labels are sorted in ascending order. Then, the sample is divided into M bins. Finally, the accuracy and average confidence are computed on each bin and reported on the diagram. If the accuracy equals the average confidence (i.e. the points are on the diagonal line) the model is considered to be perfectly calibrated. However, this representation can be misleading since there exists a trade-off between the number of bins chosen and the confidence estimated. Indeed, an insufficient number of samples in some bins can result in an inefficient estimation. Even with a large number of observations, it is not always obvious to assert the quality of the calibration. Figure~\ref{fig:RelDiag} illustrates this point by showing a reliability diagram with different values of $M$. The left plot shows the reliability diagram with 10 bins while the right plot uses 50 bins. 

Both plots are using the same data. The left plot would make us believe that the calibration is almost perfect for the Platt method, whereas it is not so obvious in the right plot. 

Synthetic measures of the calibration based on the reliability diagram also allow judging the quality of calibrated probabilities. Recently, the Expected Calibration Error (ECE) has been proposed to summarize the calibration benefits~\cite{CalECE}. This measure is computed as the average of the difference between the accuracy and the confidence over the bins: $$\text{ECE} = \sum_{m=1}^{M}\frac{|B_m|}{N}|\acc(B_m) - \conf(B_m)|,$$ where $N$ is the total number of data points. Alternatively, one can also want to measure the worst possible gap between the ideal calibration and the confidence estimated which is possible through the Maximum Calibration Error (MCE)~\cite{CalECE}. ECE and MCE suffer from the same limitations as the reliability diagram~\cite{MeasureCal}. For instance, the distribution of predicted probabilities is often left-skewed. This has the effect to put more weight on certain bins densely populated which affects the ECE. Therefore, choosing the number of bins amounts to a bias-variance tradeoff. Indeed, when $M$ is large, the number of points in each bins becomes small leading to an increase of the variance while the bias tends to decrease. 

Because of the above limitations, we decide to use another classical way to evaluate the calibration. It consists in using the full distribution of the calibrated probabilities and comparing them to the true output. This is the purpose of the Binary Cross-Entropy (BCE) and the Brier Score (BS)~\cite{Brier}. The BCE and the BS have been computed as follows: $$\text{BCE} = -\frac{1}{N}\sum_{i=1}^{N} y_i\log(\widetilde{p_i}) + (1-y_i)\log{(1-\widetilde{p_i})},
$$
$$\text{BS} = \frac{1}{N}\sum_{i=1}^{N}(y_i-\widetilde{p_i})^{2}.
$$

As we can see, the BS is similar to the MSE often used to measure the error in regression problems. In this particular setting, it can be interpreted as the mean prediction error. The BCE comes from the specification of the likelihood in binary classification problems as stated in section~\ref{Plattscaling}. It is also used as a loss function for training binary classifiers. Additionally, we compute the Mean Default Rate (MDR) for each method: $$\text{MDR}= \left(\frac{1}{N}\sum_{i=1}^{N}\widetilde{p}_i \right) \times 100.
$$
This last measure is useful to control the resulting bias of the calibrated probabilities. Indeed, a perfectly calibrated classifier should produce probabilities that have the same mean default rate as the true observed default rate. 

\section{Empirical results} \label{results}

In this section, we report the results obtained for the logistic regression and the neural network on the different sets of features. The models are optimized by minimizing the BCE and the optimal threshold $\tau^{\star}$ is chosen to maximize the F1 score. Both models are evaluated using the F1 score and we also report the precision and the recall to diagnose the possible model weaknesses and the ROC-AUC and PR-AUC to spot the effect of imbalanced learning. We then move on to the calibration results. We compare the results of the Platt method to our SURE method. 

\subsection{Comparison of models}

For both the logistic regression and the feed forward neural network, we use the same learning and evaluation procedure. First, the model is trained on the learning set and at the same time evaluated on the validation set. We implement the early stopping rule such that the model stops learning when validation performance starts to diverge according to the performance on the training set. Once the stopping rule is reached, the model's parameters are saved~\cite{Prechelt98}. The remaining model is then used on the test set to assess its performance on unseen data. 


As we can see in both Table~\ref{table1} and Table~\ref{table2}, it seems that the two models do not suffer from overfitting. Indeed, for all different sets of features, the training and testing results are very close and even better when looking at the logistic regression testing results in Table~\ref{table1}. Thus, we can infer that our early stopping procedure has been successful to prevent overfitting and both models can generalize well with the data provided. As stated in Section~\ref{evaluationMetric}, the AUC-ROC is not a good choice for assessing the performance of our models. Indeed, since we are in an imbalanced class problem, it produces good results for both models because it does not focus only on the positive class. The AUC-PR and F1 score are more reliable indicators of the model performance for predicting consumer default. On average, the observed gap between the AUC-ROC and the F1 score is 0.2, which is quite high. On the feature side, an interesting thing to remark is that the set of dynamic features provides better results than the static ones. Indeed, the testing F1 scores (i.e. results obtained on unseen data) are improved by around 0.10 if we consider the logistic regression and by around 0.15 considering the feed forward neural network. Thus we can infer that dynamic data contains more informative features for customer default prediction. 

\begin{table*}[htbp]
\ra{1.3}
\begin{adjustbox}{max width=\textwidth}
\begin{tabular}{@{}lccccccccccc@{}}
\toprule
& \multicolumn{3}{c}{Static features} & \phantom{abc}& \multicolumn{3}{c}{Dynamic features} &
\phantom{abc} & \multicolumn{3}{c}{All features}\\
\cmidrule{2-4} \cmidrule{6-8} \cmidrule{10-12}
& Training & Validation & Testing && Training & Validation & Testing && Training & Validation & Testing \\ \midrule
F1 & 0.40 & 0.40 & 0.40 && 0.51 & 0.53 & 0.53 && 0.51 & 0.52 & 0.53 \\ 
Recall & 0.70 & 0.70 & 0.71 && 0.48 & 0.51 & 0.52 && 0.48 & 0.51 & 0.51 \\ 
Precision & 0.27 & 0.28 & 0.28 && 0.54 & 0.54 & 0.55 & & 0.54 & 0.54 & 0.55 \\
AUC-ROC & 0.62 & 0.63 & 0.63 && 0.71 & 0.73 & 0.73 && 0.72 & 0.74 & 0.73 \\ 
AUC-PR & 0.30 & 0.32 & 0.31 && 0.49 & 0.50 & 0.51 && 0.50 & 0.51 & 0.51 \\ 
Loss & 0.23 & 0.23 & 0.23 && 0.21 & 0.21 & 0.21 && 0.21 & 0.21 & 0.21 \\
\bottomrule
\end{tabular}
\end{adjustbox}
\caption{Logistic regression performance on the different sets.}
\label{table1}
\end{table*}

\begin{table*}[htbp]
\ra{1.3}
\begin{adjustbox}{max width=\textwidth}
\begin{tabular}{@{}lccccccccccc@{}}
\toprule
& \multicolumn{3}{c}{Static features} & \phantom{abc}& \multicolumn{3}{c}{Dynamic features} &
\phantom{abc} & \multicolumn{3}{c}{All features}\\
\cmidrule{2-4} \cmidrule{6-8} \cmidrule{10-12}
& Training & Validation & Testing && Training & Validation & Testing && Training & Validation & Testing \\ \midrule
F1 & 0.40 & 0.38 & 0.39 & & 0.56 & 0.53 & 0.54 & & 0.56 & 0.52 & 0.55 \\ 
Recall & 0.82 & 0.81 & 0.81 && 0.62 & 0.53 & 0.55 && 0.61 & 0.51 & 0.54 \\ 
Precision & 0.27 & 0.25 & 0.25 && 0.53 & 0.53 & 0.53 && 0.53 & 0.53 & 0.55 \\
AUC-ROC & 0.59 & 0.60 & 0.60 && 0.77 & 0.77 & 0.77 && 0.78 & 0.76 & 0.77 \\ 
AUC-PR & 0.28 & 0.30 & 0.28 && 0.54 & 0.54 & 0.53 && 0.54 & 0.53 & 0.53 \\ 
Loss & 0.23 & 0.23 & 0.23 && 0.20 & 0.20 & 0.20 && 0.19 & 0.20 & 0.20 \\
\bottomrule
\end{tabular}
\end{adjustbox}
\caption{Neural network performance on the different sets.}
\label{table2}
\end{table*} 

We can also notice that the feed forward neural net seems to be unstable on static data since its F1 score, AUC-ROC, AUC-PR values are worse than for the logistic regression. We suspect that the difference is due to both the set of features and the model complexity. Indeed, on one hand, the feed forward neural network learns complex data interactions on the static features which appear to be uninformative for the given task. On the other hand, the logistic regression can be seen as a feed forward neural network without hidden layers, thus leading to a simpler model with fewer parameters. The resulting performance of the logistic regression is better because this model doesn't learn these complex interactions. This result can be connected to the adversarial attack notion which exists in computer vision~\cite{Szegedy,Goodfellow}. The general idea is to fool the model by introducing a little perturbation noise into the input data which doesn't change the image perception for a human. This perturbation causes the deep learning model to produce false predictions. 

When we investigate the results for the dynamic and the whole feature sets, we can observe that the feed forward neural network reaches better testing results than the logistic regression. This improvement comes from the model architecture. With the hidden layers being fully connected to each other, the network learns deeper representations of the inputs by automatically creating feature interactions. These representations allow the network to learn more complex patterns than the logistic regression, which explains why the results are improved. Even if there is an undeniable improvement of the performance, we can notice that results are very close between Table~\ref{table1} and Table~\ref{table2}. For example, the F1 score is superior by only 0.02 for the neural network which is not high. Here, we want to stress the fact that appropriate measures should be used to evaluate the model depending on the use case. For the same dataset, several studies have shown that a neural network is a more accurate model while our results suggest only little improvement~\cite{creditscoring1,Imtiaz}. 

\subsection{Evaluation of the calibration}

We present the results obtained by the different calibration techniques. We used the same procedure for training and testing as standard papers on calibration. First, we fit the model on the predicted probabilities of the validation set and use the learned parameters to calibrate the predicted probabilities of the test set~\cite{Platt99,tempscalingNN,FieldCal}. The measures realized on the test set are unbiased estimates of the calibration error. 
The first observation that can be done about the results gathered in Tables~\ref{tab:table3} and~\ref{tab:table4} is that uncalibrated probabilities do not reflect the true probabilities. Indeed, the mean probability goes from 45.26\% for the Neural network to 47.04\% for the logistic regression whereas the true empirical default rate is 22.13\%. This emphasizes the need for probability calibration.  

\begin{table*}[htbp]
\ra{1.3}
\begin{adjustbox}{max width=\textwidth}
\begin{tabular}{@{}lcccccccc@{}}
\toprule
& \multicolumn{2}{c}{Mean default rate} & \phantom{abc}& \multicolumn{2}{c}{BCE} &
\phantom{abc} & \multicolumn{2}{c}{BS}\\
\cmidrule{2-3} \cmidrule{5-6} \cmidrule{8-9}
& Validation & Testing && Validation & Testing && Validation & Testing \\ \midrule
 Uncalibrated              & 46.73 & 47.04 && 0.610 & 0.613 && 0.210 & 0.211 \\

Platt                      & 22.13 & 22.45 && 0.460 & 0.459 && 0.144 & 0.143 \\

SURE (sigmoid)             & 24.82 & 25.10 && 0.464 & 0.464 && 0.146 & 0.145 \\

SURE (Kumaraswamy)         & 24.81 & 25.10 && 0.467 & 0.468 && 0.147 & 0.146 \\

Platt + SURE (sigmoid)         & 22.41 & 22.72 && 0.455 & 0.452 && 0.143 & 0.141 \\

SURE (sigmoid) + Platt         & 22.13 & 22.47 && 0.454 & 0.452 && 0.142 & 0.141 \\

Platt + SURE (Kumaraswamy) & 21.95 & 22.26 && 0.460 & 0.459 && 0.144 & 0.143 \\

SURE (Kumaraswamy) + Platt & 22.13 & 22.47 && 0.456 & 0.454 && 0.143 & 0.141 \\
\bottomrule
\end{tabular}
\end{adjustbox}
\caption{Logistic regression calibration.}
\label{tab:table3}
\end{table*}
 
 \begin{table*}[htbp]
\ra{1.3}
\begin{adjustbox}{max width=\textwidth}
\begin{tabular}{@{}lcccccccc@{}}
\toprule
& \multicolumn{2}{c}{Mean default rate} & \phantom{abc}& \multicolumn{2}{c}{BCE} &
\phantom{abc} & \multicolumn{2}{c}{BS}\\
\cmidrule{2-3} \cmidrule{5-6} \cmidrule{8-9}
& Validation & Testing && Validation & Testing && Validation & Testing \\ \midrule
Uncalibrated              & 45.26 & 45.48 && 0.580 & 0.584 && 0.194 & 0.194 \\

Platt                     & 22.13 & 22.37 && 0.437 & 0.436 && 0.137 & 0.136 \\

SURE (sigmoid)            & 24.69 & 24.93 && 0.440 & 0.439 && 0.138 & 0.137 \\

SURE (Kumaraswamy)        & 24.70 & 24.96 && 0.441 & 0.442 && 0.138 & 0.138 \\

Platt + SURE (sigmoid)    & 22.39 & 22.69 && 0.438 & 0.436 && 0.137 & 0.136 \\

SURE (sigmoid) + Platt    & 22.13 & 22.43 && 0.438 & 0.435 && 0.137 & 0.136 \\

Platt + SURE (Kumarswamy) & 21.81 & 22.05 && 0.437 & 0.436 && 0.137 & 0.136 \\

SURE (Kumarswamy) + Platt & 22.13 & 22.45 && 0.438 & 0.436 && 0.137 & 0.137 \\
\bottomrule
\end{tabular}
\end{adjustbox}
\caption{Neural network calibration.}
\label{tab:table4}
\end{table*}

Regarding the performance of the various calibration methods, we report that the Platt and SURE methods lead to a sharp improvement of uncalibrated probabilities.
When we investigate both methods separately, Platt scaling appears to give slightly better results than SURE. Nevertheless, this new approach achieves promising results. When we combine both methods that is we feed calibrated probabilities by one method to the other method, we notice an improvement of the BCE and BS measures for the probabilities predicted by logistic regression. We also emphasize that there is little variance in the performance of the stacking models which indicates the stability of the method.

\section{Conclusion}

In this paper, we have investigated classification models for determining the probability that a consumer will default on its credit card. The obtained results give us some guidelines for this kind of prediction problem. First, the class imbalance should be identified in order to choose an appropriate measure of performance. In the case of a severe class imbalance, it is recommended to use a measure that relies on positive class detection, such as the F1 score, to have an unbiased estimate of the model's true performance. Secondly, temporal features, if available, must be added to the model as they can significantly improve its learning ability and performance. Depending on the frequency of the data, it could be interesting to perform pre-processing steps in order to extract useful information such as descriptive statistics of that series. This has been left for further studies. On the contrary, one should take care of uninformative features which can lead to false data representations and decreased performance when learning with a complex model such as a neural network. Thirdly, we stress out the fact that real-world applications should output predicted probabilities that reflect the true unobservable probabilities in order to diminish uncertainty. Our results show that it can be done accurately and efficiently by using a staking of the Platt and SURE calibration methods.

\printbibliography

\appendix

\section{Derivatives of the sigmoid function}\label{ap:sigderive}

\begin{align*}
G_{\theta}(\widehat{p_i}) &= \frac{1}{1+e^{\theta_1 \widehat{p_i}+\theta_2}}\\
\frac{\partial G_{\theta}(\widehat{p_i})}{\partial \widehat{p_i}} &=
 \theta_{1}G_{\theta}(\widehat{p_i})(1-G_{\theta}(\widehat{p_i}))\\
\frac{\partial G_{\theta}(\widehat{p_i})}{\partial \theta_1} &= 
 \widehat{p_{i}}G_{\theta}(\widehat{p_i})(1-G_{\theta}(\widehat{p_i}))\\
\frac{\partial G_{\theta}(\widehat{p_i})}{\partial \theta_2} &= 
 G_{\theta}(\widehat{p_i})(1-G_{\theta}(\widehat{p_i}))\\
\frac{\partial G_{\theta}(\widehat{p_i})}{\partial \widehat{p_i} \partial \theta_1} &= 
 G_{\theta}(\widehat{p_i})(1-G_{\theta}(\widehat{p_i})) + \theta_1 \widehat{p_i}G_{\theta}(\widehat{p_i})(1-G_{\theta}(\widehat{p_i}))(1-2G_{\theta}(\widehat{p_i})) \\
 \frac{\partial G_{\theta}(\widehat{p_i})}{\partial \widehat{p_i} \partial \theta_2} &= 
  \theta_1G_{\theta}(\widehat{p_i})(1-G_{\theta}(\widehat{p_i}))(1-2G_{\theta}(\widehat{p_i}))
\end{align*}

\section{Derivatives of the Kumaraswamy cumulative distribution}\label{ap:Kumderive}

\begin{align*}
G_{\theta}(\widehat{p_i}) &= 1-(1-\widehat{p_i}^{\theta_1})^{\theta_2} \\
\frac{\partial G_{\theta}(\widehat{p_i})}{\partial \widehat{p_i}} &= \theta_1\theta_2 \widehat{p_i}^{\theta_1 -1}(1-\widehat{p_i}^{\theta_1})^{\theta_2 -1}\\
\frac{\partial G_{\theta}(\widehat{p_i})}{\partial \theta_1} &= \theta_2 \log{(\widehat{p_i})} \widehat{p_i}^{\theta_1}(1-\widehat{p_i}^{\theta_1})^{\theta_2-1}\\
\frac{\partial G_{\theta}(\widehat{p_i})}{\partial \theta_2} &= -\log{(1-\widehat{p_i}^{\theta_1})}(1-\widehat{p_i}^{\theta_1})^{\theta_2}\\
\frac{\partial G_{\theta}(\widehat{p_i})}{\partial \widehat{p_i} \theta_1} &= \theta_2(\widehat{p_i}^{\theta_1 -1}(1-\widehat{p_i}^{\theta_1})^{\theta_2-1}\left(1 + \theta_1 \log{(\widehat{p_i})} \right) - \theta_1\widehat{p_i}^{2\theta_1 -1}(\theta_2-1)(1-\widehat{p_i}^{\theta_1})^{\theta_2-2}\log{(\widehat{p_i})})\\
\frac{\partial G_{\theta}(\widehat{p_i})}{\partial \widehat{p_i} \theta_2} &= \theta_1\widehat{p_i}^{\theta_1 -1}(1-\widehat{p_i}^{\theta_1})^{\theta_2-1}(1 +\theta_2\log{(1-\widehat{p_i}^{\theta_1})})
\end{align*}

\end{document}